\documentclass[fleqn,10pt]{wlscirep}
\usepackage[utf8]{inputenc}
\usepackage[T1]{fontenc}
\usepackage{cleveref}
\usepackage{multicol}
\usepackage{multirow}
\usepackage{makecell}
\usepackage{float}
\usepackage{subcaption}
\usepackage{hyperref}

\title{Using Smartphones to Study Vaccination Decisions in the Wild}

\author[1,2,*]{Nicolò Alessandro Girardini}
\author[13]{Arkadiusz Stopczynski}
\author[9,10,11]{Olga Baranov}
\author[7,8]{Cornelia Betsch}
\author[12]{Dirk Brockmann}
\author[3,4,*]{Sune Lehmann}
\author[3,5,6,*]{Robert Böhm}

\affil[1]{Department of Information Engineering and Computer Science (DISI), University of Trento, Italy}
\affil[2]{Mobile and Social Computing Lab (MobS), Fondazione Bruno Kessler (FBK), Trento, Italy}
\affil[3]{Copenhagen Center for Social Data Science (SODAS), University of Copenhagen, Denmark}
\affil[4]{Department of Applied Mathematics and Computer Science, Technical University of Denmark, Lyngby, Denmark}
\affil[5]{Faculty of Psychology, University of Vienna, Austria}
\affil[6]{Department of Psychology, University of Copenhagen, Denmark}
\affil[7]{Health Communication, Implementation Science, Bernhard Nocht Institute for Tropical Medicine, Germany}
\affil[8]{Health Communication, Institute for Planetary Health Behaviour, University of Erfurt, Germany}
\affil[9]{Fraunhofer Institute for Translational Medicine and Pharmacology ITMP, Immunology, Infection and Pandemic Research, Munich, Germany}
\affil[10]{German Centre for Infection Research (DZIF), partner site Munich, Germany}
\affil[11]{Division of Infectious Diseases and Tropical Medicine, LMU University Hospital, LMU Munich, Germany}
\affil[12]{Center Synergy of Systems AND Center for Interdisciplinary Digital Sciences AND Faculty of Biology AND Faculty of Physics, TUD Dresden University of Technology, Dresden, Germany}
\affil[13]{Independent Researcher}

\affil[*]{Corresponding authors: ngirardini@fbk.eu, sljo@dtu.dk, robert.boehm@univie.ac.at}


\begin{abstract}
One of the most important tools available to limit the spread and impact of infectious diseases is vaccination. It is therefore important to understand what factors determine people's vaccination decisions. To this end, previous behavioural research made use of, (i) controlled but often abstract or hypothetical studies (e.g., vignettes) or, (ii) realistic but typically less flexible studies that make it difficult to understand individual decision processes (e.g., clinical trials). Combining the best of these approaches, we propose integrating real-world Bluetooth contacts via smartphones in several rounds of a game scenario, as a novel methodology to study vaccination decisions and disease spread.
In our 12-week proof-of-concept study conducted with $N$ = 494 students, we found that participants strongly responded to some of the information provided to them during or after each decision round, particularly those related to their individual health outcomes. In contrast, information related to others' decisions and outcomes (e.g., the number of vaccinated or infected individuals) appeared to be less important. We discuss the potential of this novel method and point to fruitful areas for future research.
\end{abstract}
\begin{document}

\flushbottom
\maketitle

\thispagestyle{empty}

\section{Introduction}
\label{sec:intro}

Vaccination is an effective and safe prevention measure to protect people from infectious diseases, saving 2-3 million lives worldwide every year \cite{2018gvap}. 
In the special case of the COVID-19 pandemic, it has been estimated that vaccines prevented about 20 million deaths during the first year of COVID-19 vaccination only \cite{watson2022global}. 
Despite this success, vaccine hesitancy---i.e., delay in acceptance or refusal of vaccination despite the availability of vaccination services---poses a major threat to global health and addressing it is therefore a priority area for national and international health organisations \cite{who10threats, macdonald2015vaccine, dube2013vaccine}. 
Accordingly, building an understanding of why people decide in favor or against vaccination is a crucial challenge for our times.

For recommended vaccinations, the personal benefits (e.g., reduced risk of infection) are expected to exceed the potential costs (e.g., risk of vaccine-adverse events). 
In addition, most vaccines also provide social benefits due to herd immunity \cite{fine2011herd}. 
That is, as a population's vaccination rate increases, the spreading of the pathogen is reduced, such that non-vaccinated individuals are indirectly protected. 
Therefore, someone's vaccination may also benefit others who either cannot get vaccinated (e.g., infants, immunocompromised persons) or do not want to get vaccinated (e.g., vaccine-hesitant persons).
Considering both its personal and social benefits, vaccination poses a strategic interaction, in which individuals have to decide whether to prosocially benefit non-vaccinated others (accountable or unaccountable) by getting vaccinated or whether to free ride on the indirect protection provided by vaccinated others, resulting in a social dilemma \cite{bauch2004vaccination, betsch2013inviting, bohm2022prosocial}. 

Given the potential role of personal and social benefits, the vaccination decision should depend on both (i) internal states, such as personal attitudes, perceptions and expectations, and experiences related to the disease and to the vaccine, as well as (ii) external data, such as information about others' vaccination behaviour, vaccine uptake in a population, or the incidence of diseases \cite{brewer2017increasing, betsch2015using, brewer2007meta, dube2013vaccine}. 
Regarding the latter, previous research has indeed shown that people consider both, the decisions of others, and the welfare consequences for others when making their own vaccination decisions. 
However, the evidence regarding the direction of such effects is mixed. 
Some studies indicate that informing people about the social benefits of vaccination increases their vaccination intentions \cite{betsch2013inviting, betsch2017benefits, logan2018have, jung2021concerns}. 
In contrast, other studies provided evidence that free riding may occur when people learn that they are protected via herd immunity \cite{bohm2016selfish, ibuka2014free, chapman2012using}. 
Moreover, further studies showed that people increase their vaccination intentions when they believe that others are (planning to get) vaccinated, too, indicating adherence to perceived social norms \cite{hershey1994roles, abdallah2021social, schmelz2021overcoming}.

\subsection{Spatial and Temporal Dynamics in Vaccination Decisions and Their Consequences}
\label{subsec:intro_dynamics}

The mixed results regarding the impact of social information and its consequences on individual vaccination decisions may be partly attributed to the heterogeneity in methods used in prior research.
That is, depending on the specific research question, different studies have increased the salience of certain contextual cues, emphasizing some information (e.g., social norms \cite{brewer2017increasing, abdallah2021social}, social benefits and its beneficiaries \cite{bohm2022prosocial, betsch2017benefits}, varying personal benefits depending on the amount of vaccinated others \cite{betsch2013inviting,jung2021concerns}) while omitting others.
Such differences in the salience of certain cues emerge easily, because when laboratory and online studies use artificial or imagined vaccination decisions, they ignore the spatial and temporal dynamics of vaccination decisions and their potential consequences as they occur in reality. 
That is, first, vaccination decisions in studies are typically independent from these dynamics, but real diseases are transmitted via some form of (direct or indirect) physical contact between individuals (i.e., spatial dynamics). 
Specifically, transmissions through direct contact can only take place when, at the time of contact, one individual is infectious and the other one is susceptible. 
Second, a vaccination decision temporally precedes its consequences, while personal and social information, such as experiences from own or others' previous vaccinations, may be available and updated over time (i.e., temporal dynamics).

Neglecting these spatial and temporal dynamics of vaccination decisions and their consequences is a major limitation of prior research's external validity. 
An obvious solution is to study actual vaccination decisions and their consequences `in the wild.' 
However, a severe drawback of this approach is that the decision processes, such as what kind of personal and social information is available to people in their decision making, are often unavailable to researchers. 
Here, we aim to overcome these limitations by presenting a novel experimental design that combines the best of both approaches, that is, studying the determinants and consequences of vaccination decisions while also considering the spatial and temporal dynamics as they unfold in the real world.

\subsection{Using Smartphones to Study Vaccination Decisions and Infection Spread}
\label{subsec:intro_phones}

We use an established interactive vaccination game that builds on a SIR (Susceptible-Infectious-Recovered) model of disease dynamics, thus capturing the dynamic personal and social benefits of vaccinations depending on the population's vaccination rate \cite{bohm2016selfish}. 
Various versions of this interactive vaccination game (I-Vax) have been successfully applied to investigate vaccination behaviour and potential behavioural interventions, with games played several rounds having usually taken place in controlled lab-settings and one-shot versions as online studies \cite{betsch2016detrimental, korn2017drawbacks, bohm2017behavioural, korn2018social, bohm2019willingness, meier2020individual, korn2020vaccination}. 
The key advantage of this gamified method is that decisions are not hypothetical, but they have actual consequences by providing behaviour-contingent monetary incentives, thus reducing the intention-behaviour gap \cite{sheeran2016intention}. 

Here, we go a step further by implementing the I-Vax game in a natural scenario, taking into account the spatial and temporal dynamics of vaccination decisions and their consequences.
We give players augmented information on the natural environment through the game mechanics, which are based on the real-world experiences of the participants. 
In our approach, we study a cohort of nearly 500 individuals, all freshmen at a Danish university, where infections take place via Bluetooth contact between participants' smartphones---that is, infections can only happen when two participating individuals are physically close one another (infection range up to ten meters, but with a higher probability of infections if the distance is lower \cite{sekara2014strength}).
This models the spatial properties of real-world disease spreading, i.e., infections happen because of actual physical contact between infectious and susceptible individuals. 
Through the implemented application the participants could take the decision to vaccinate and protect themselves from the fictional disease.
Our I-Vax study, took place across an entire 13-week semester of students attending a single campus, thus generating millions of contacts \cite{stopczynski2014measuring}.

We implemented 12 independent game rounds, each lasting one week, where players had their naturally occurring and repeated physical contacts with each other, to model the temporal properties of vaccination decisions and their consequences. 
Across the 12 rounds, we varied the information available to players ranging from no information in the beginning of the game, via local information about the vaccination status of neighbors, to full population-level information in the final rounds of the game (see \Cref{subsec:intro_set-up} below for full details).
As each information setting was deployed multiple times, participants could learn from previous rounds and potentially change their perceptions of the disease and vaccination as well as their beliefs about other participants' behaviours over time.
This mirrors repeated vaccination decisions, such as in the case of influenza or COVID-19 (i.e., booster doses). As such, our set-up allows a `time travel' through a repeated vaccination history under behaviour-contingent incentives that would otherwise require years of observations `in the wild.' 
Importantly, given that all I-Vax decisions to vaccinate or not were made via a smartphone application, we have access to all the information that participants received during the game, and can therefore examine the impact of all pieces of information on players' vaccination decisions.

In summary, in this study we combine the benefits of maximum experimental control with a highly valid setting: the study's setup allows controlling for or tracking numerous factors related to decision makers' personal and social benefits of vaccination as in artificial, but highly controlled vaccination studies. At the same time the set-up is externally valid due to the mimicking of spatial and temporal dynamics of disease spreading in natural conditions. Thus, this results in a controlled yet natural vaccination scenario.

\subsection{Study Set-Up and Data Handling}
\label{subsec:intro_set-up}

Our exploratory proof-of-concept experimental study was conducted with $N$ = 494 students at a Danish university as part of the SensibleDTU project \cite{stopczynski2014measuring}. 
Participants were given a smartphone which recorded sensor data and usage data.
This allowed the implementation of the I-Vax game, including the use of Bluetooth sensors for disease transmission.
In this section, we describe an overview of the basic set-up and the detailed factors that were exogenously varied during the experimental study. 
We provide more details on the parametrisation and implementation in the Methods section, \Cref{sec:methods}. 

\paragraph{Game Mechanics.} The game mechanics were designed to recreate the vaccination process and the spreading of the disease to be as similar as possible to natural conditions. 
Some participants were infected at the start of each round and thus treated as `patients zero', or `seeds'.
These players then spread the disease to physically close susceptible players, who were neither vaccinated, infected nor recovered.
The infection underwent a phase of simulated incubation of the disease.
Thus, players did not become infectious immediately following the infection, but only after a fixed amount of time. They also recovered from the disease after a fixed amount of time. 
Players could decide to get vaccinated at any point in time during the round.
As under natural conditions, the vaccination was not immediately effective, but it became so after eight hours.
In our implementation, the vaccine is 100\% effective.

In each round, players started with 100 points. Points are deducted to simulate fixed vaccination costs (e.g., the effort to make an appointment with the doctor and the time spent receiving the jab), vaccination side effects, and the effects of getting sick with the disease when unvaccinated.
Importantly, whereas vaccination costs are low, fixed and certain, the probability and severity of vaccination side-effects follow an exogenous probability function, and disease costs are determined by the endogenous probability of having a (Bluetooth) contact with an infectious player coupled with an exogenous severity function (for a detailed rationalisation, see \cite{bohm2016selfish}).
Moreover, the average point loss due to encountering the disease outweighs the side-effects vaccination loss.
Finally, to motivate players to make rational decisions, we used a point system to track individual-level performance, which determined rewards at the end of the experiment---the players with the most points after the full 12 rounds semester won iPads and FitBits.

\paragraph{Manipulations.} We vary the information environment in a within-participants design with a fixed order. 
All information is available to participants through their smartphone app. 
In rounds 1-4, players only rely on their personal attitude and experience in the game (\textit{no feedback} condition). In rounds 5-8, players receive information about their surrounding environment (\textit{local feedback} condition). 
This means that every day they learn about the amount of infected, vaccinated, and susceptible players among the players they interacted with on the previous day. 
Finally, in rounds 9-12, instead of receiving the statistics of their contacts, players receive information about the whole population (\textit{global feedback} condition). At the end of each round, players also get the cumulative information for the whole round, in both the local and global feedback conditions.
We define these as the different conditions of the \textit{environmental feedback} we give them.

With these different information environments, we can explore whether the social information available in the spatial environment affects participants' vaccination decisions. 
We modeled the individual vaccination decisions using the information cues available to the players (e.g., share of infected, vaccinated, and susceptible players) as factors.
Therefore, we are able to distinguish between different information cues (e.g., share of infected, vaccinated, and susceptible players) and information effects as suggested by previous laboratory and online research (e.g., prosocial vaccination, free riding, social-norm following). 
The realistic set-up, which considers the spatial and temporal dynamics of natural vaccination decisions and infections, allows for the estimation of these effects without sacrificing data availability and experimental control.

\paragraph{Analytical Approach.}
As a first step, we analyzed players' vaccine uptake across rounds, only considering the individual experience they gained by playing consecutive rounds. 
This implies drawing on information of their performance at the end of each round, i.e., whether there was a point loss and whether it was due to potential side effects in case of vaccination or due to potential infection with the disease in case of non-vaccination.
As a second step, we added the environmental information that were given to players in rounds 5-12. The analysis rests on different feedback about other players' behaviours and outcomes at the end of each round.
The final step consisted of adding the time component (day) interior to each weekly round, to see how the effects of different factors change when considering a more dynamic setting, including information received within each round.

For the first two steps, \textit{Mixed Linear Effects (MLE)} models \cite{raudenbush2002hierarchical} have been used for the analyses, considering each round as a `box' and, thus, reflecting the analysis that is more commonly done in laboratory and online experiments and even in many real-world clinical trials. 
For the final step, \textit{Survival Models} \cite{survival_book} were used, as they were the optimal method to promote the benefit of our experiments' temporal and spatial resolution within each round.

This study is meant as a proof of concept for the deployment of virtual epidemics and their application for studying vaccination behaviors. This pilot study suffers from some drawbacks, most notably some rounds of disease dying off due to pure chance and the declining engagement over the time course of the study (see \Cref{sec:discussion} for more details on these issues).
Therefore, we do not make any claims about descriptive (mean level) differences in vaccine uptake, but we rather focus on individual factors of vaccine uptake. All the analyses should be treated as exploratory. We provide a full description of the specific methods underlying the presented results in the Methods section, \Cref{sec:methods}.

\section{Results}
\label{sec:res}

\subsection{Players' Behaviour given Individual Outcome in Previous Round}
\label{subsec:ind_out}

\begin{table}[t]
    \centering
    \caption{MLE model predicting vaccinations decisions across all rounds by feedback of personal consequences after each round ($k$ = 3,934 observations by $n$ = 469 players).}
    \begin{tabular}{l|c|c|c|c|c|c}
        \multirow{2}{*}{Factor} & \multicolumn{2}{c|}{Coefficient} & \multicolumn{2}{c|}{Odds Ratio} & \multicolumn{2}{c}{z-value} \\ \cline{2-7}
        & Est. & Std.Err. & Val. & 95\% CI & z-value & P($>|$z$|$) \\
        \hline
        Intercept & -1.63 & 0.21 & 0.20 & [0.13,0.29] & -7.803 & 6.04e-15 *** \\
        Points Loss & 1.90 & 0.16 & 6.70 & [4.91,9.15] & 11.992 & $<$ 2e-16 *** \\
        Previous Vaccination Decision & 2.13 & 0.16 & 8.41 & [6.13,11.54] & 13.191 & $<$ 2e-16 *** \\
        Round Number & -0.19 & 0.05 & 0.83 & [0.75,0.91] & -4.018 & 5.87e-05 *** \\
        Local Feedback & -0.23 & 0.13 & 0.80 & [0.62,1.02] & -1.807 & 0.071 \\
        Global Feedback & -0.44 & 0.14 & 0.64 & [0.49,0.84] & -3.211 & 0.001 ** \\
        Points Loss: Previous Vaccination Decision & -2.02 & 0.23 & 0.13 & [0.09,0.21] & -8.962 & $<$ 2e-16 *** \\
        \hline
        \multicolumn{7}{c}{} \\
        \multicolumn{7}{c}{Significance level: ‘***’ 0.001 ‘**’ 0.01 ‘*’ 0.05}
    \end{tabular}
    \label{tab:individual}
\end{table}

As described in \Cref{subsec:intro_set-up}, the first analysis involves predicting vaccination decisions across rounds by using individual outcomes in the previous round. 
The results are shown in \Cref{tab:individual}. 
The model was constructed so that the intercept represents a baseline setting, where players only received individual feedback, they did not have a point loss (neither from the disease nor a vaccination side effect), and they did not get vaccinated in the previous round. 
Finding a negative coefficient shows that, overall, the decision to avoid vaccination was more common than getting vaccinated. 
The other factors are dummy variables that represent the three different feedback conditions (no feedback, local feedback, global feedback) described above and whether or not there was a point loss. 
We included an interaction term between point loss and the previous decision, which allowed us to control whether the point loss came from the disease or from vaccination side effects.

\begin{figure}[t]
    \centering
    \includegraphics[width=15cm]{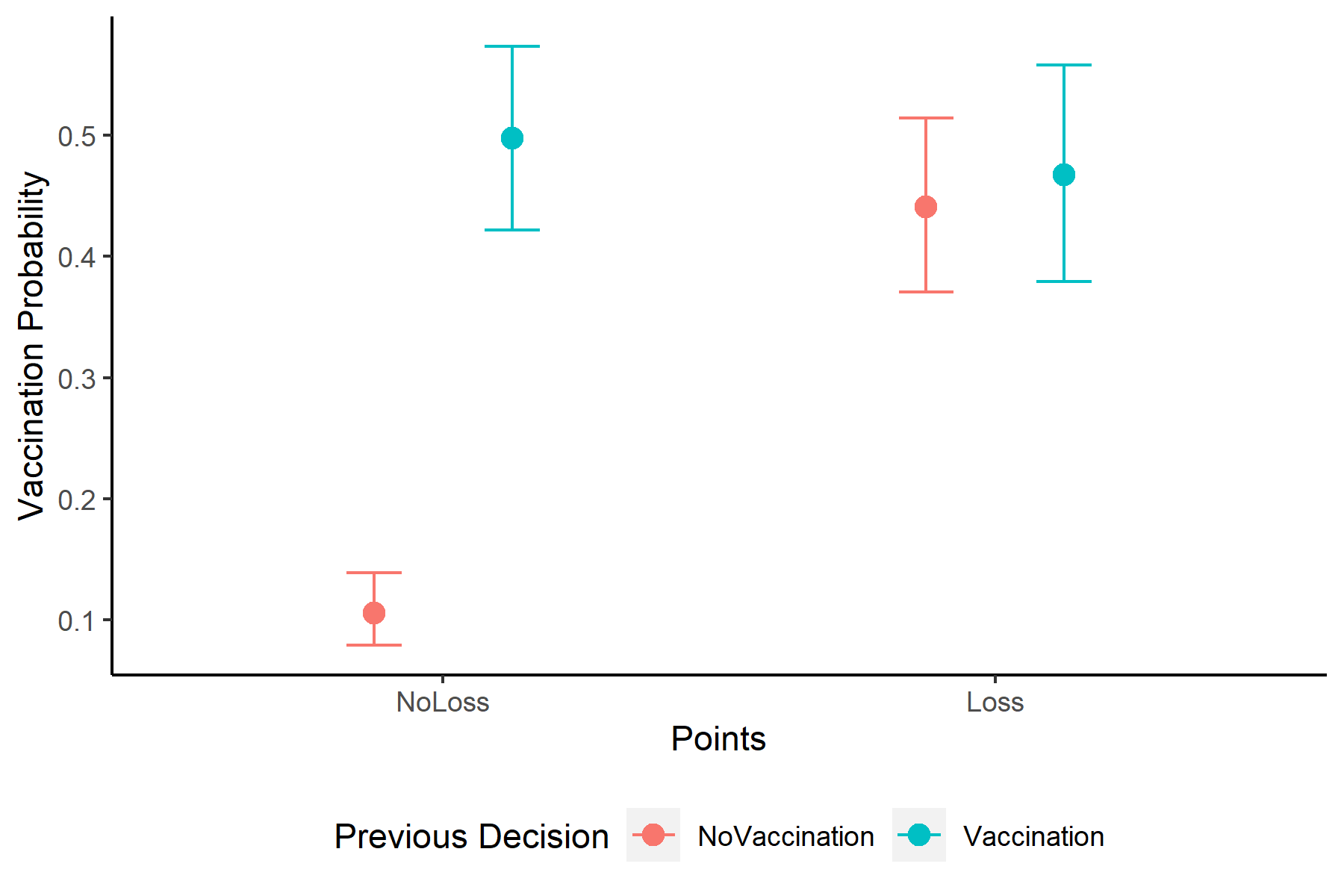}
    \caption{\textbf{Vaccination Probability given Point Loss in Former Round and Previous Decision.} Effects of point loss in the former round and the previous decision on the probability to get vaccinated in the subsequent round ($k$ = 3,934 observations by $n$ = 469 players). Dots represent the estimated probability and error bars the standard error. We see that a point loss implies a lower vaccination probability, but only for players that did not vaccinate in the previous round.}
    \label{fig:points_decprev}
\end{figure}

We found positive effects of \textit{Point Loss} and \textit{Previous Vaccination Decision}, qualified by a significant interaction effect. 
As shown in \Cref{fig:points_decprev}, players who did get vaccinated in the previous round were equally likely to get vaccinated in the next round, irrespective of whether they experienced vaccination side effects or not. 
In contrast, players who did not get vaccinated in the previous round were more likely to get vaccinated subsequently when they got sick (and lost points) than when they did not experience disease costs and thus were successful free-riders. 

This result indicates that players responded in a  rational way to negative outcomes, probably to maximize their outcomes. 
Interestingly, players appear to be more sensitive to the costs of infection than to the costs of vaccination. 
The analysis shows further (negative) effects of \textit{Round Number} and \textit{Global Feedback}, indicating that players were less likely to get vaccinated in later rounds, and that providing global feedback about others' decisions and outcomes also reduced their own likelihood to get vaccinated.
Taken together, our findings suggest that players were sensitive to the outcomes of the game, providing a first proof-of-concept for the study setup's validity.

\begin{table}[H]
    \centering
    \caption{MLE model predicting vaccinations decisions across rounds 5-12 by feedback of personal consequences and environmental feedback after each round ($k$ = 2,481 observations by $n$ = 427 players).}
    \begin{tabular}{l|c|c|c|c|c|c}
        \multirow{2}{*}{Factor} & \multicolumn{2}{c|}{Coefficient} & \multicolumn{2}{c|}{Odds Ratio} & \multicolumn{2}{c}{z-value} \\ \cline{2-7}
        & Est. & Std.Err. & Val. & 95\% CI & z-value & P($>|$z$|$) \\
        \hline
        Intercept & -2.35 & 0.54 & 0.10 & [0.03,0.28] & -4.323 & 1.54e-05 *** \\
        Points Loss & 1.96 & 0.25 & 7.08 & [4.35,11.53] & 7.872 & 3.50e-15 *** \\
        Previous Vaccination Decision  & 2.60 & 0.30 & 13.49 & [7.48,24.32] & 8.655 & $<$ 2e-16 *** \\
        Round Number & -0.24 & 0.10 & 0.79 & [0.65,0.96] & -2.381 & 0.017 * \\
        Global Feedback & -0.02 & 0.73 & 0.98 & [0.23,4.12] & -0.031 & 0.976 \\
        Points Loss: Previous Decision Vaccination & -2.05 & 0.34 & 0.13 & [0.07,0.25] & -5.985 & 2.17e-09 *** \\
        Vaccinated EndOfRound & -0.08 & 0.63 & 0.92 & [0.27,3.17] & -0.130 & 0.897 \\
        Infected EndOfRound & 1.03 & 2.57 & 2.80 & [0.02,430.58] & 0.400 & 0.689 \\
        Susceptible EndOfRound & 0.63 & 0.48 & 1.88 & [0.73,4.83] & 1.305 & 0.192 \\
        Global Feedback: Vaccinated EndOfRound & -0.28 & 1.10 & 0.76 & [0.09,6.58] & -0.251 & 0.802 \\
        Global Feedback: Infected EndOfRound & -0.63 & 2.87 & 0.53 & [0.001,148.67] & -0.219 & 0.827 \\
        Global Feedback: Susceptible EndOfRound & -0.31 & 0.82 & 0.73 & [0.15,3.65] & -0.381 & 0.703 \\
        \hline
        \multicolumn{7}{c}{} \\
        \multicolumn{7}{c}{Significance level: ‘***’ 0.001 ‘**’ 0.01 ‘*’ 0.05}
    \end{tabular}
    \label{tab:env_feed}
\end{table}

\subsection{Players' Behaviour Given Environmental Feedback}
\label{subsec:env_feedback}

\begin{table}[t]
    \centering
    \caption{Survival Model predicting whether a vaccination happens before the end of the round across rounds 5-12 by feedback of personal consequences and environmental feedback after each round ($k$ = 2,003 observations by $n$ = 382 players).}
    \begin{tabular}{l|c|c|c|c|c|c}
        \multirow{2}{*}{Factor} & \multicolumn{3}{c|}{Coefficient} & Hazard Ratio & \multicolumn{2}{c}{$\chi^2$ Test} \\ \cline{2-7}
        & Est. & Std.Err. & 95\% CI & Exp(coef) & $\chi^2$ & p-value \\
        \hline
        Points Loss & 0.4652 & 0.3293 & [-0.1803,1.1107] & 1.592 & 2.00 & 0.16 \\
        Previous Decision Vaccination & 1.4514 & 0.2394 & [0.9823,1.9206] & 4.269 & 36.77 & 1.3e-09 *** \\
        Round Number & -0.0812 & 0.0957 & [-0.2686,0.1063] & 0.922 & 0.72 & 0.40 \\
        Global Feedback & -18.6673 & 4.9075 & [-28.2859,-9.0487] & 7.81e-09 & 14.47 & 1.4e-04 *** \\
        \makecell[l]{Points Loss: \\ \hspace{.5cm}Previous Decision Vaccination} & -0.3220 & 0.4625 & [-1.2285,0.5846] & 0.725 & 0.48 & 0.49 \\
        Daily Vaccinated & 0.0013 & 0.0052 & [-0.0089,0.0116] & 1.001 & 0.06 & 0.80 \\
        Daily Infected & 0.0132 & 0.0252 & [-0.0362,0.0626] & 1.013 & 0.27 & 0.60 \\
        Daily Susceptible & 0.0167 & 0.0031 & [0.0107,0.0228] & 1.017 & 29.13 & 6.8e-08 *** \\
        Global Feedback: Daily Vaccinated & -0.0674 & 0.1184 & [-0.2995,0.1648] & 0.935 & 0.32 & 0.57 \\
        Global Feedback: Daily Infected & 0.4425 & 0.2354 & [-0.0189,0.9039] & 1.557 & 3.53 & 0.06 \\
        Global Feedback: Daily Susceptible & 0.2625 & 0.0595 & [0.1459,0.3790] & 1.300 & 19.48 & 1.0e-05 *** \\
        \hline
        \multicolumn{7}{c}{} \\
        \multicolumn{7}{c}{Significance level: ‘***’ 0.001 ‘**’ 0.01 ‘*’ 0.05}
    \end{tabular}
    \label{tab:survival}
\end{table}

To analyze the impact of the environmental feedback on players' vaccination decisions, we distinguish between the following pieces of information shown to each player at the end of each round in the local and global feedback conditions: share of vaccinated players (\textit{Vaccinated EndOfRound}), share of infected players (\textit{Infected EndOfRound}) and share of susceptible players (\textit{Susceptible EndOfRound}). 
The baseline setting is now the local feedback condition and the main effects of the feedback factors refer to their influence in this condition only. 
Therefore, we also added interaction terms between the information players received and the feedback condition, capturing the impact of information in the global feedback condition.

As shown in \Cref{tab:env_feed}, the feedback on other players' behaviours and outcomes did not significantly affect their own vaccination decision, in any of the feedback conditions. 
In contrast, the information on personal outcomes remains to significantly affect players' decisions, as reported above. 
This is surprising, considering how players behaved differently following the feedback settings (as demonstrated in \Cref{subsec:ind_out}). 
Our conclusion in this regard is that participants, in our experiment, value much more the individual feedback rather than the environmental feedback---another interpretation could be a lack of interest in the game over time, which we cannot rule out due to the fixed order of the feedback conditions.

\subsection{Players' Behaviour Given Daily Feedback}
\label{subsec:daily_feedback}

\begin{figure}[t]
    \centering
    \includegraphics[width=15cm]{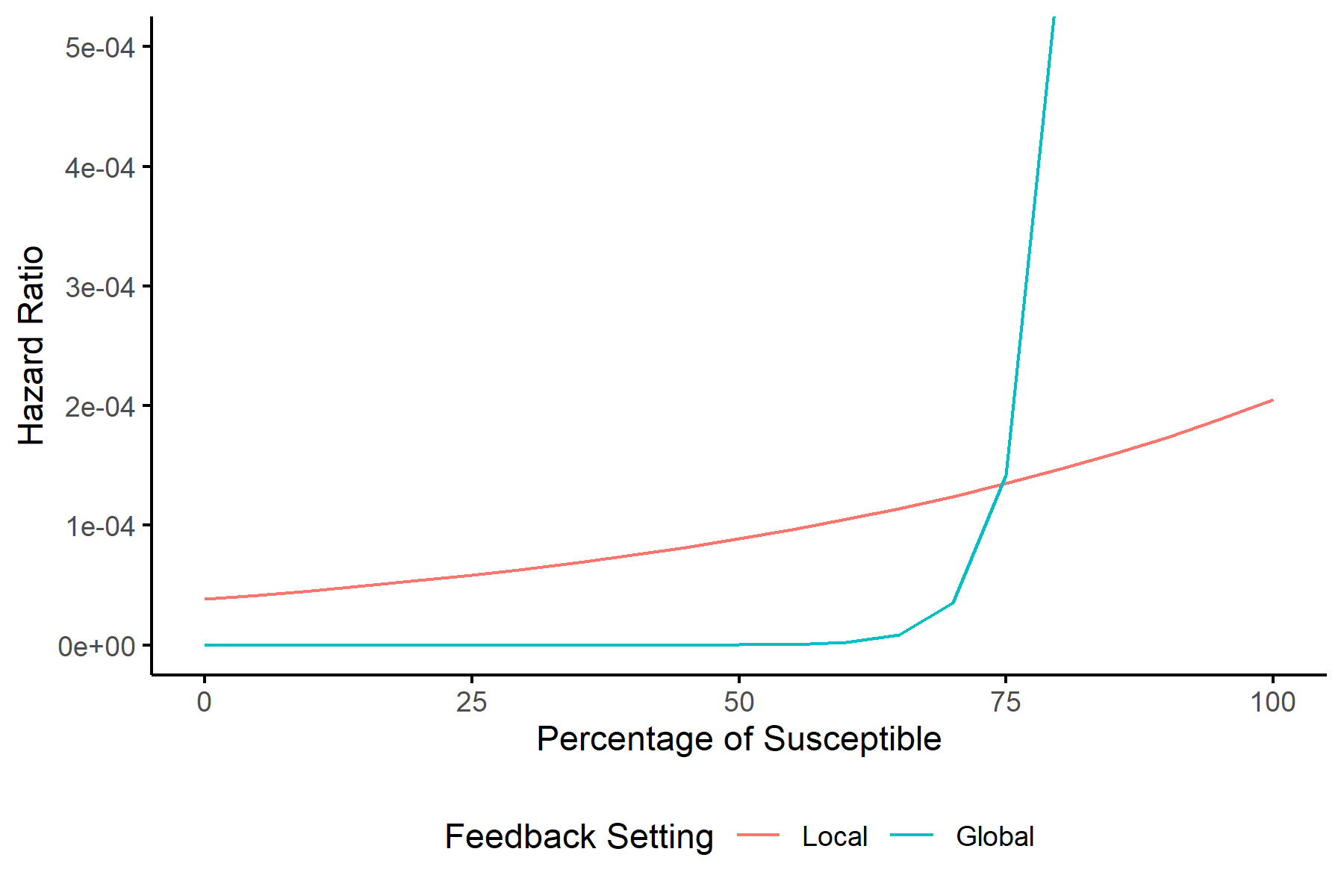}
    \label{fig:daily_susc}
    \caption{\textbf{Likelihood of Vaccination given Daily Susceptible Feedback.} Hazard Ratio (i.e., likelihood of vaccinating) of daily feedback about the share of susceptible others in the local vs. global feedback condition ($k$ = 2,003 observations by $n$ = 382 players). We see that, for the \textit{Global Feedback} condition, the likelihood of vaccinating exponentially increases when the number of susceptible individuals is high.}
    \label{fig:daily}
\end{figure}

The final part of the analysis regards the most unique part of our methodology, which adds the time dimension within each round. 
As in a natural setting, players received daily feedback on whether other players had been vaccinated, infected, or susceptible. 
Specifically, at the start of each day, they receive information, relative to the previous day, about their environment and either on other players they have been in contact with (local feedback condition) or on all other players in the population (global feedback condition). 
We use this information shown to participants to predict their likelihood of getting vaccinated within a Survival Analysis model \cite{survival_book}, with the Hazard Ratio indicating the (increased or decreased) likelihood of getting vaccinated over time. 
Given the structure of the daily feedback, the analysis is using the players' behavior starting from day two and lasting for the remaining 6 days of the round.

Our first observation is that 84.4\% of the vaccinations over all rounds happened during the first day of the week. 
This may indicate that players were strongly affected by the feedback they received at the end of the rounds (as analysed above), rather than the feedback they received during the progress of the rounds. 
The survival model is estimated on the remaining share of vaccination decisions that were not made on the very first day of a round.

The specific model is Cox Proportional Hazard (CoxPH) \cite{survival_book}, which, similarly to MLE, explains the impact of each factor on vaccination decisions. 
The results are shown in \Cref{tab:survival}. 
As in the previous analysis, the baseline setting is the local feedback condition, that is, feedback about other players' decisions and outcomes with whom one had physical interactions as assessed through Bluetooth contacts. 
There is a positive effect of the \textit{Previous Vaccination Decision}. 
This indicates that even those players who did not get vaccinated on the very first day of a new round were more likely to get vaccinated later during the round when they were already vaccinated in the previous one. 
The main effects of the \textit{Global Feedback} condition and the daily feedback on the number of susceptible players, i.e. \textit{Daily Susceptible}, were qualified by a significant interaction effect between these variables. 
As shown in \Cref{fig:daily}, players were more likely to get vaccinated when they received daily feedback about an increasing share of susceptible others, but with a bigger difference in the global setting than in the local setting.
This appears reasonable, given that if many people are at risk, the urge to protect oneself could increase. 
In absolute numbers, in the global setting, the likelihood of getting vaccinated really has a spike when about 75\% or more of the players in the general population were susceptible, reaching an Hazard Ratio of $\approx 0.15$.

\section{Discussion}
\label{sec:discussion}

The main goal of the present research was to devise and gain seminal experience with a novel method to study vaccination decisions and their consequences in an externally valid yet high-control set-up. 
Whereas previous experimental research on vaccination decisions typically ignored the spatial and temporal dynamics of vaccination and disease spread, we implemented a scenario in which people may contract a fictional disease by Bluetooth contact with other players in the real world. 
Specifically, the use of a smartphone application allowed the spreading of the disease to occur in physical space, following time patterns for the events (vaccination, incubation time, etc.) that reflect real-world mechanics. 
To make the decisions and outcomes meaningful, we implemented an interactive vaccination game with behaviour-contingent incentives \cite{bohm2016selfish}.

To validate this novel approach, we conducted a comprehensive experimental investigation involving $N$ = 494 participants over a span of 12 weeks.
Throughout this period, which was divided into rounds in a gamified fashion, we manipulated the information given to participants regarding other players' choices on vaccination and their consequences.
Overall, our results suggest that players were sensitive to the feedback and associated outcomes. 
For instance, we found that players increased their likelihood of getting vaccinated when they got infected in the previous round. Moreover, we found some evidence of players reacting to other others' behaviors, too. Particularly, players who did not decide immediately to get vaccinated, were later more likely to get vaccinated when they saw higher numbers of susceptible players in subsequent rounds.
This could indicate that players perceived either a potential danger that the infection could easily reach them, or some prosocial motivation to protect others with one's own vaccination \cite{bohm2022prosocial}. 
Taken together, both selfish rationales, as well as prosocial motives, could account for the observed behaviours.

The feedback manipulation was a novel aspect of the experimental case study with our I-Vax game.
We found some differences in players' responses to feedback from interactions with players vs. feedback about the whole study population. For example, participants valued more their own experience and did not consider much the feedback given at the end of the round, be that global or local. They, instead, valued highly the daily information about the number of players that were still susceptible.
This observation deserves attention in future research. 
Although we implemented this variation mainly for exploratory reasons, future studies could investigate the role of social norms, free-riding motivation, and prosocial concern when people learn about a more local, and potentially psychologically closer, group versus a more distant but larger group. 
Clearly, differences in how people value such information could inform policymakers to improve vaccination communication.

Another interesting effect to investigate would be the effect of individual attitudes towards vaccinations in our experiment. We conjecture that feedback might have a bigger impact on players with a rather low or medium vaccination attitude, as those with a high vaccination attitude would vaccinate nonetheless. In fact, we see that most of the vaccinations happen during the first day of the round, indicating how important the individual attitudes might be.

\subsection{Limitations}
\label{subsec:limit}

Although the method utilized here is powerful, the experimental set-up had some limitations that should be considered when interpreting our findings and planning future research endeavors.
First, in some instances of the local feedback condition, a scarcity of initially infected seed users may have created a misleading perception of reduced disease transmission, potentially leading to lower vaccine uptake in subsequent rounds.
Fortunately, this issue had a limited impact on our study, as vaccinations showed a consistent decline over time, rather than a sudden drop after these rounds.
Nevertheless, we recommend cautious seed user selection in future implementations, for example, conducting synthetic runs to test the spread of the disease at the start of the rounds.

Second, we acknowledge that certain parameter conditions in our set-up, such as 100\% vaccine efficacy, a 0.3 probability of vaccination side effects, and uniform risk categorization among users, do not mirror real-world scenarios.
However, the parametrization of I-Vax game's set-up, allows for easy adjustment of these parameters.

Third, in our set-up we consecutively implement three different conditions, in which players have different environmental information to consider for their vaccination decision.
This influence was evident in the decreasing vaccination rates over time.
For future implementations, introducing user groups subjected to different conditions simultaneously could enhance result robustness.

Involving many participants for a study such as this can be demanding in terms of recruiting users. This can still be facilitated by the widespread presence of smartphones, onto which users could just install the application.
While studying a population well apt with digital tools, such as university students, can help in this issue, it still has the drawback of not being representative of the whole population. 
Furthermore, concerns might arise regarding users' data protection, which we suggest to handle with attention (also see \Cref{subsec:comms}).

Finally, proper to the gamification approach, dropout can be a concern.
Fortunately, our study experienced relatively low dropout, potentially due to economic incentives driving player engagement.
Nevertheless, the inconsistent participant numbers across rounds introduced some variability, which we have mitigated through the adaptation of the applied models.

\subsection{Conclusion}
\label{subsec:conclusion}

In summary, our novel method combines the best of highly controlled laboratory and online experiments, and large-scale clinical trials studying actual vaccine uptake. We propose that running experiments using this method could provide important insights before rolling out said clinical trials (e.g. to evaluate interventions aiming at increasing vaccine uptake), and potentially reduce the attitude-behaviour gap that plagues implementation research.

\section{Methods}
\label{sec:methods}

The first proof-of-concept implementation of our novel method was integrated into the \textit{SensibleDTU} experiment \cite{stopczynski2014measuring}. SensibleDTU is a living lab experiment conducted at the Technical University of Denmark (DTU), in which 1000 students were provided a smartphone, and sensor data was collected, along with surveys and app usage information.

In this setting, the I-Vax game was implemented as a smartphone application. 
This implementation had several advantages. 
Firstly, the modelling of the disease spreading followed real-life mechanics. We set the $R_0$ parameter to 3, to mimic the spread of seasonal influenza \cite{influenza_r0} (meaning that an infected person on average spreads the disease to 3 others). The spreading follows physical contact between the participants, using  Bluetooth technology \cite{sekara2014strength}.

Gamification is another fundamental aspect of our novel methodology, as the participants are incentivised to behave responsively to the game with economic incentives, based on an in-game point system. This, in conjunction with the time and space components of the game, reconstructed a natural scenario, both for the participants' experience and for disease transmission.

\subsection{I-Vax Game Implementation}
\label{subsec:game}

The modelling of the game, as described above, is composed of both game design and disease modelling. In the following paragraphs, we will describe in detail our implementation.

\paragraph{Point System} A point system was devised to guide players' decisions and make them understand the outcome of the game and evaluate their actual performance. Players started each round with 100 fitness points. Following specific events, they could lose points, simulating real-life consequences and costs:

\begin{itemize}
    \item \textbf{Vaccination:} Real vaccinations often require spending time and effort (e.g., to book the appointment and to actually get the injection), and they may also cost actual money. To simulate this, players had to pay 10 points to get vaccinated. The vaccine was 100\% effective.
    \item \textbf{Vaccination side effects:} It is possible to have allergic reactions or other side effects after a vaccination. To simulate these events, a vaccinated player had a 0.3 probability of getting side effects and losing points: this probability allows to control for rare severe consequences and common very mild symptoms like arm soreness \cite{nichol1996side}. To account for this, the point loss could range from 0 to 90 points (on top of the vaccination cost), with an average of 30 points lost.
    \item \textbf{Infection:} Different symptoms are possible in real-world diseases, but the consequences of getting infected are usually worse than possible vaccination side effects. Thus, when infected, the point loss still ranges in the 0-90 points interval, but with a higher average of 50 points lost. Low values of point loss are included to simulate asymptomatic infections or those with very mild symptoms.
\end{itemize}

\begin{figure}[t]
    \centering
    \includegraphics[width=\linewidth]{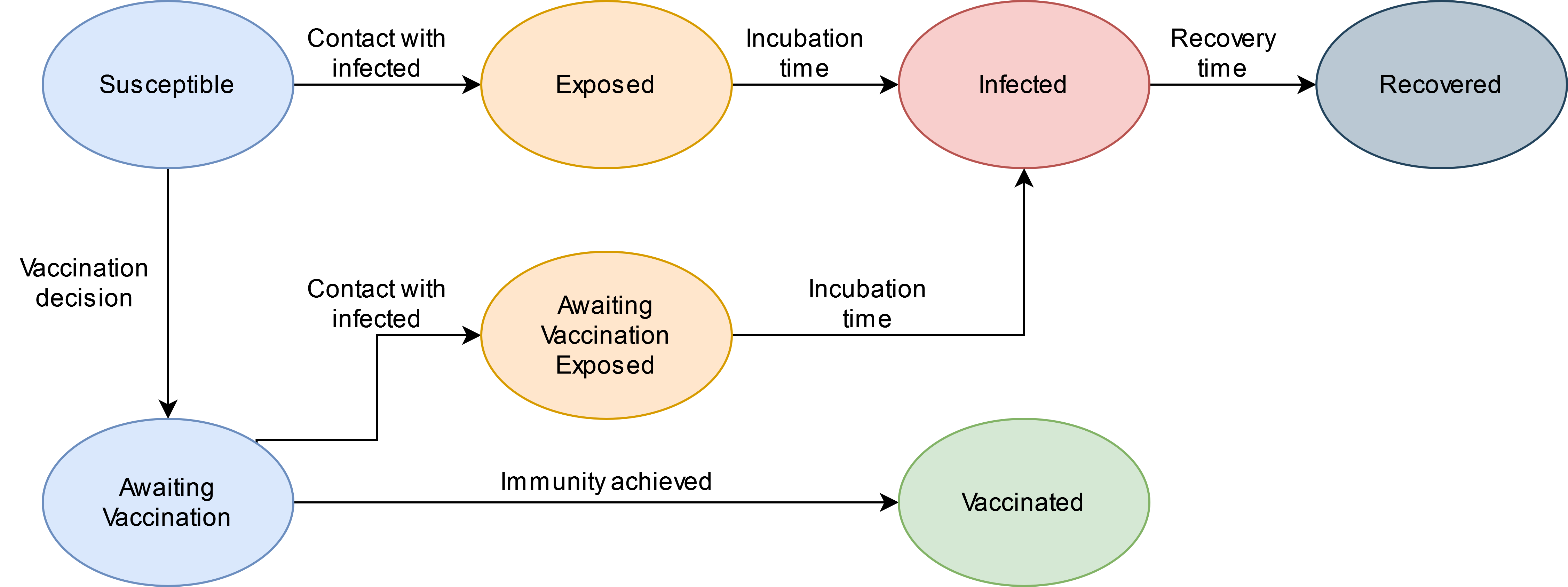}
    \caption{\textbf{Finite State Automaton of the Experiment.} All players start from the susceptible state and they can remain so, decide to get vaccinated or get infected and then recover. Labels on the transitions show what event is triggering them in the partly probabilistic-causal game framework.}
    \label{fig:fsa}
\end{figure}

\paragraph{Finite State Automaton} To model the states of the players during the game and the transitions between the states, we constructed a finite state automaton, represented in \Cref{fig:fsa}. The essential states are: \textit{Susceptible (S), Vaccinated (V), Infected (I) and Recovered (R)}. Other states (\textit{Awaiting vaccination (A), Awaiting vaccination Exposed (AE) and Exposed (E)}) were introduced to model specific mechanics and to manage the state transitions. At the start of a round, every player is susceptible. They can either remain so until the end of the round (if they do not get vaccinated or infected), they can get vaccinated to become immune to the disease, or they can get infected when being close to someone who is already infected (i.e., he is infectious). Players who got infected, heal from the disease after a set amount of time, becoming recovered (and immune).

\paragraph{Vaccination Mechanics} The core mechanics are vaccination and infection, which are sets of transitions. The vaccination is the only transition that players can trigger with their own decision. The vaccine is not effective immediately, as it is in real life, the body has to respond to it and register its information. In the same way, in our implementation, the vaccine is effective (at 100\% effectiveness) after ca. 8 hours. The player has to pay the 10 points immediately (as he would spend time and money to have the vaccine shot, before achieving immunity). To model the time interval when the vaccine is not yet effective, a state was added, i.e., awaiting vaccination (A). During this time, the player was not protected from the virus and could get infected, transitioning to the state: awaiting vaccination exposed (AE), which then leads to an infection. Only after the immunity was achieved the player could experience side-effects, finally reaching state V.

\paragraph{Infection Mechanics} To start the spread of the disease, 10\% of the players were chosen as initial seeds and their state was set to infected (I). This simulates the fact that at the start of an epidemic, there are no externally imposed containment efforts. This strategy also gives control over the initialization of the experiment. The seeds, and the newly infected, could spread the disease to players in state S or A, so that the R0 index (i.e. how many players get infected on average by a single infected) was equal to 3, simulating a seasonal influenza \cite{influenza_r0}. We also simulate the incubation time, in which the person exposed to the disease is not yet infected: for ca. 6 hours, the player was in state exposed (E). After the incubation time, the player was in infected state I, now being infectious themselves, for ca. 18 hours. After this period of time, players would recover, being unable to get infected again (or to vaccinate), getting to state R.

\paragraph{Feedback Settings} The final mechanics of the game are repeated measures, i.e., multiple rounds were played, simulating the seasonality of the disease, like in a real-world scenario. Twelve rounds (or waves) were played. We changed the feedback given to players (through the I-Vax application) every 4 rounds, allowing the study of different feedback conditions. In this way, we achieve the controllability possible in artificial experiments in our realistic setting. In our experimental setting, we controlled for information about the players' surrounding environment, and implemented 3 treatment conditions:

\begin{itemize}
    \item \textbf{Rounds 1-4:} At the end of each of these rounds, we gave players individual-level information, consisting of the outcome of the round, i.e., remaining points, final state and final decision. We studied the impact of this information on the next round's vaccination decision. This condition is called \textit{No Feedback}, because of the absence of environmental feedback. This type of information was provided throughout all the conditions. 
    \item \textbf{Rounds 5-8:} The feedback given to players involves their local environment, i.e. the players they had contact with. This is called \textit{Local Feedback} condition. Every day (starting from day 2 of the week) players were given information about the players they had interacted with the day before, in the form of the percentage of infected, vaccinated or susceptible among them. The same information was given at the end of the round, aggregating the daily feedback.
    \item \textbf{Rounds 9-12:} Similar to the previous condition, players receive daily and weekly feedback about the percentage of infected, vaccinated or susceptible, but relative to the whole population of players instead of their contacts. This is called \textit{Global Feedback} condition.
\end{itemize}

\subsection{Communication with Participants and Privacy Considerations}
\label{subsec:comms}

A core part of the experiment is the communication with the participants. There are many possible types of bias that can alter the outcome of an experiment depending on how this communication happens. One specific worrisome bias, when it comes to communication, is response bias \cite{bogner2016response, podsakoff2003common}. This describes the fact that participants may take certain actions just because they are part of the experiment or they are influenced by the communication with the researchers. To counter this kind of bias, in our experiment, the communication with players has been done solely through the use of the I-Vax application implemented for the game. In the application, players could find the rules of the game and all the feedback information they were given, either daily or at the end of the rounds. Moreover, they were also made aware of the effect of herd immunity in the game, i.e., with 70\% of the players vaccinated the probability of the disease spreading would have been zero. In this way, it was possible to simulate how people receive and perceive information in a real-world scenario and to understand how they are affected and which information affects them the most.

The privacy of the participants of the game was strictly respected. They were assigned an identifier based on their SensibleDTU ID. Thus, for further information relative to these concerns we refer the reader to \cite{stopczynski2014privacy, sapiezynski2015tracking}. In the data we share, these IDs have been further anonymised.

\subsection{Analysis Methods}
\label{subsec:an_methods}

Following the structure of our experimental setting, we identified as fitting analysis two models, i.e., \textit{Mixed Linear Effects (MLE)} \cite{raudenbush2002hierarchical} and \textit{Survival Analysis} \cite{survival_book}. The analysis part is important to validate in practice the validity of our method. The data has been normalized and centered before the analysis.

MLE models are one of the standards for analyzing data involving human behaviour. They consist of linear regression, controlling for different factors (fixed effects) while also allowing the grouping of observations (i.e., random effects) where there could be a dependency or common characteristics or, as in our case, repeated measures. Each player had to make repeated decisions in different rounds, thus being the random effects. In this way, we established the independence between observations pertaining to different players, being able to assess individual performances and outcomes. The fixed effects, instead, include all the information that was given to players (remaining points, status, previous decisions, feedback), so to see their effect on the participants' decisions. We use the logit link function, as the outcome is a binary decision, which maps the output of the model into the interval [0,1], resulting in the vaccination probability. These models have been used to map the rounds of the experiment, as it is not really possible to capture the flow of time inside them. Following this, we used as input the snapshots of the end of the rounds, by feeding the models the individual outcome and the weekly feedback given to players.

\Cref{eq:model1} and \Cref{eq:model2} describe more in detail the models described in \Cref{subsec:ind_out} and \Cref{subsec:env_feedback} respectively. Here the models predict the \textit{decision} of the players, grouping the observations by player with the random effect $(1|user)$, with the main effects and the interactions effects (indicated by the column symbol).

\begin{equation}
    \label{eq:model1}
    decision \sim (1|user)+points+decision\_prev+wave\_no\_en+feedback\_setting+points:decision\_prev
\end{equation}

\begin{equation}
    \label{eq:model2}
    \begin{aligned}
        decision \sim &(1|user)+points+decision\_prev+wave\_no\_en+feedback\_setting+points:decision\_prev+ \\
        & vaccinated\_eor+infected\_eor+susceptible\_eor+feedback\_setting:vaccinated\_eor+ \\
        & feedback\_setting:infected\_eor+feedback\_setting:susceptible\_eor
    \end{aligned}
\end{equation}

For the analysis of the daily feedback, instead, a more time-dependent model was needed. Survival Analysis models time until a set event happens at time t (usually the event is death, hence the name survival). Using days as the time measure, allowed us to analyze what happens during a round and to use as factors the daily feedback that was given to players. Since the scope of this analysis was to look into which of the factors were significant towards the vaccination decision, rather than modelling how many days would pass before a participant decided to get vaccinated, the specific model chosen was \textit{Cox Proportional Hazards (CoxPH)} \cite{survival_book}. This model identifies how much different factors contribute towards the event, thus, in our case, the vaccination. Finally, its interpretation is very similar to the MLE models, making the whole analysis comprehensive and coherent at the same time.

The following \Cref{eq:model3} describes the survival model used in \Cref{subsec:daily_feedback}. Similarly to above, it groups the users and uses both main and interaction effects of the factors we consider. The output is the final \textit{decision} along with the \textit{day} in which that decision would be taken (i.e., the time in which the event has happened). Note that we did not consider this in our results, as we consider most important the factors that foster the decision rather than its timing.

\begin{equation}
    \label{eq:model3}
    \begin{aligned}
        Surv(day,decision) \sim &(1|user)+points+decision\_prev+wave\_no\_en+feedback\_setting+points:decision\_prev+ \\
        &vaccinated\_surv+infected\_surv+susceptible\_surv+feedback\_setting:vaccinated\_surv+ \\
        &feedback\_setting:infected\_surv+feedback\_setting:susceptible\_surv
    \end{aligned}
\end{equation}

\section*{Author contributions statement}

S.L. and R.B. supervised the project. S.L., A.S., D.B., R.B., C.B. designed the experiment and the main implementation was done by A.S.. The data was prepared by N.A.G., A.S., O.B. and analyzed by N.A.G.. The paper was written by N.A.G. and R.B., and all the authors provided comment and revised and approved the final draft.

\section*{Acknowledgements}
S.L. is thankful to the Villum Foundation for the Young Investigator Grant that made this work possible. The authors thank Radu G\v{a}tej for his work on preparing the data.

\section*{Ethical Statement}
The research project and data collection was registered with and approved by the Danish Data Supervision Authority before data collection commenced. All data was collected with informed consent and with every participant able to withdraw from the study and have their data deleted. This protocol, implemented in 2012 and 2013, was in effect similar to the rules being introduced with the EU General Data Protection Regulation (GDPR) which came into effect in May 2018. In the present release, to comply with GDPR, the data has been stripped of personally identifying information and the data has been reduced in such a way that there is no reasonable likelihood of re-identification occurring.

\section*{Competing interests}

All authors declare no financial or non-financial competing interests.

\section*{Data and Code Availability}

Both the anonymized data and the code used for the analysis are available in a GitHub repository at the following \href{https://github.com/Svidon/I-Vax-Code}{link}.

\bibliography{biblio}

\begin{thebibliography}{10}
\urlstyle{rm}
\expandafter\ifx\csname url\endcsname\relax
  \def\url#1{\texttt{#1}}\fi
\expandafter\ifx\csname urlprefix\endcsname\relax\def\urlprefix{URL }\fi
\expandafter\ifx\csname doiprefix\endcsname\relax\def\doiprefix{DOI: }\fi
\providecommand{\bibinfo}[2]{#2}
\providecommand{\eprint}[2][]{\url{#2}}

\bibitem{2018gvap}
\bibinfo{author}{{World Health Organization (WHO)}}.
\newblock \bibinfo{title}{{2018 assessment report of the Global Vaccine Action Plan: strategic advisory group of experts on immunization}}.
\newblock \bibinfo{howpublished}{\emph{Technical documents} \url{https://apps.who.int/iris/handle/10665/276967}} (\bibinfo{year}{2018}).
\newblock \bibinfo{note}{Last accessed: 14-08-2023}.

\bibitem{watson2022global}
\bibinfo{author}{Watson, O.~J.} \emph{et~al.}
\newblock \bibinfo{journal}{\bibinfo{title}{{Global impact of the first year of COVID-19 vaccination: a mathematical modelling study}}}.
\newblock {\emph{\JournalTitle{The Lancet Infectious Diseases}}} \textbf{\bibinfo{volume}{22}}, \bibinfo{pages}{1293--1302}, \doiprefix\url{https://doi.org/10.1016/S1473-3099(22)00320-6} (\bibinfo{year}{2022}).

\bibitem{who10threats}
\bibinfo{author}{{World Health Organization (WHO)}}.
\newblock \bibinfo{title}{{Ten threats to global health in 2019}}.
\newblock \bibinfo{howpublished}{\emph{Online} \url{https://www.who.int/news-room/spotlight/ten-threats-to-global-health-in-2019}} (\bibinfo{year}{2019}).
\newblock \bibinfo{note}{Last accessed: 14-08-2023}.

\bibitem{macdonald2015vaccine}
\bibinfo{author}{MacDonald, N.~E.} \emph{et~al.}
\newblock \bibinfo{journal}{\bibinfo{title}{{Vaccine hesitancy: Definition, scope and determinants}}}.
\newblock {\emph{\JournalTitle{Vaccine}}} \textbf{\bibinfo{volume}{33}}, \bibinfo{pages}{4161--4164}, \doiprefix\url{https://doi.org/10.1016/j.vaccine.2015.04.036} (\bibinfo{year}{2015}).

\bibitem{dube2013vaccine}
\bibinfo{author}{Dub{\'e}, E.} \emph{et~al.}
\newblock \bibinfo{journal}{\bibinfo{title}{Vaccine hesitancy: an overview}}.
\newblock {\emph{\JournalTitle{Human vaccines \& immunotherapeutics}}} \textbf{\bibinfo{volume}{9}}, \bibinfo{pages}{1763--1773}, \doiprefix\url{https://doi.org/10.4161/hv.24657} (\bibinfo{year}{2013}).

\bibitem{fine2011herd}
\bibinfo{author}{Fine, P.}, \bibinfo{author}{Eames, K.} \& \bibinfo{author}{Heymann, D.~L.}
\newblock \bibinfo{journal}{\bibinfo{title}{{“Herd immunity”: a rough guide}}}.
\newblock {\emph{\JournalTitle{Clinical infectious diseases}}} \textbf{\bibinfo{volume}{52}}, \bibinfo{pages}{911--916}, \doiprefix\url{https://doi.org/10.1093/cid/cir007} (\bibinfo{year}{2011}).

\bibitem{bauch2004vaccination}
\bibinfo{author}{Bauch, C.~T.} \& \bibinfo{author}{Earn, D.~J.}
\newblock \bibinfo{journal}{\bibinfo{title}{Vaccination and the theory of games}}.
\newblock {\emph{\JournalTitle{Proceedings of the National Academy of Sciences}}} \textbf{\bibinfo{volume}{101}}, \bibinfo{pages}{13391--13394}, \doiprefix\url{https://doi.org/10.1073/pnas.0403823101} (\bibinfo{year}{2004}).

\bibitem{betsch2013inviting}
\bibinfo{author}{Betsch, C.}, \bibinfo{author}{B{\"o}hm, R.} \& \bibinfo{author}{Korn, L.}
\newblock \bibinfo{journal}{\bibinfo{title}{Inviting free-riders or appealing to prosocial behavior? game-theoretical reflections on communicating herd immunity in vaccine advocacy.}}
\newblock {\emph{\JournalTitle{Health Psychology}}} \textbf{\bibinfo{volume}{32}}, \bibinfo{pages}{978}, \doiprefix\url{https://doi.org/10.1037/a0031590} (\bibinfo{year}{2013}).

\bibitem{bohm2022prosocial}
\bibinfo{author}{B{\"o}hm, R.} \& \bibinfo{author}{Betsch, C.}
\newblock \bibinfo{journal}{\bibinfo{title}{Prosocial vaccination}}.
\newblock {\emph{\JournalTitle{Current opinion in psychology}}} \textbf{\bibinfo{volume}{43}}, \bibinfo{pages}{307--311}, \doiprefix\url{https://doi.org/10.1016/j.copsyc.2021.08.010} (\bibinfo{year}{2022}).

\bibitem{brewer2017increasing}
\bibinfo{author}{Brewer, N.~T.}, \bibinfo{author}{Chapman, G.~B.}, \bibinfo{author}{Rothman, A.~J.}, \bibinfo{author}{Leask, J.} \& \bibinfo{author}{Kempe, A.}
\newblock \bibinfo{journal}{\bibinfo{title}{Increasing vaccination: putting psychological science into action}}.
\newblock {\emph{\JournalTitle{Psychological Science in the Public Interest}}} \textbf{\bibinfo{volume}{18}}, \bibinfo{pages}{149--207}, \doiprefix\url{https://doi.org/10.1177/1529100618760521} (\bibinfo{year}{2017}).

\bibitem{betsch2015using}
\bibinfo{author}{Betsch, C.}, \bibinfo{author}{B{\"o}hm, R.} \& \bibinfo{author}{Chapman, G.~B.}
\newblock \bibinfo{journal}{\bibinfo{title}{Using behavioral insights to increase vaccination policy effectiveness}}.
\newblock {\emph{\JournalTitle{Policy Insights from the Behavioral and Brain Sciences}}} \textbf{\bibinfo{volume}{2}}, \bibinfo{pages}{61--73}, \doiprefix\url{https://doi.org/10.1177/2372732215600716} (\bibinfo{year}{2015}).

\bibitem{brewer2007meta}
\bibinfo{author}{Brewer, N.~T.} \emph{et~al.}
\newblock \bibinfo{journal}{\bibinfo{title}{Meta-analysis of the relationship between risk perception and health behavior: the example of vaccination.}}
\newblock {\emph{\JournalTitle{Health psychology}}} \textbf{\bibinfo{volume}{26}}, \bibinfo{pages}{136}, \doiprefix\url{https://doi.org/10.1037/0278-6133.26.2.136} (\bibinfo{year}{2007}).

\bibitem{betsch2017benefits}
\bibinfo{author}{Betsch, C.}, \bibinfo{author}{B{\"o}hm, R.}, \bibinfo{author}{Korn, L.} \& \bibinfo{author}{Holtmann, C.}
\newblock \bibinfo{journal}{\bibinfo{title}{On the benefits of explaining herd immunity in vaccine advocacy}}.
\newblock {\emph{\JournalTitle{Nature human behaviour}}} \textbf{\bibinfo{volume}{1}}, \bibinfo{pages}{0056}, \doiprefix\url{https://doi.org/10.1038/s41562-017-0056} (\bibinfo{year}{2017}).

\bibitem{logan2018have}
\bibinfo{author}{Logan, J.} \emph{et~al.}
\newblock \bibinfo{journal}{\bibinfo{title}{{‘What have you HEARD about the HERD?’ Does education about local influenza vaccination coverage and herd immunity affect willingness to vaccinate?}}}
\newblock {\emph{\JournalTitle{Vaccine}}} \textbf{\bibinfo{volume}{36}}, \bibinfo{pages}{4118--4125}, \doiprefix\url{https://doi.org/10.1016/j.vaccine.2018.05.037} (\bibinfo{year}{2018}).

\bibitem{jung2021concerns}
\bibinfo{author}{Jung, H.} \& \bibinfo{author}{Albarrac{\'\i}n, D.}
\newblock \bibinfo{journal}{\bibinfo{title}{{Concerns for others increase the likelihood of vaccination against influenza and COVID-19 more in sparsely rather than densely populated areas}}}.
\newblock {\emph{\JournalTitle{Proceedings of the National Academy of Sciences}}} \textbf{\bibinfo{volume}{118}}, \bibinfo{pages}{e2007538118}, \doiprefix\url{https://doi.org/10.1073/pnas.2007538118} (\bibinfo{year}{2021}).

\bibitem{bohm2016selfish}
\bibinfo{author}{B{\"o}hm, R.}, \bibinfo{author}{Betsch, C.} \& \bibinfo{author}{Korn, L.}
\newblock \bibinfo{journal}{\bibinfo{title}{{Selfish-rational non-vaccination: Experimental evidence from an interactive vaccination game}}}.
\newblock {\emph{\JournalTitle{Journal of Economic Behavior \& Organization}}} \textbf{\bibinfo{volume}{131}}, \bibinfo{pages}{183--195}, \doiprefix\url{https://doi.org/10.1016/j.jebo.2015.11.008} (\bibinfo{year}{2016}).

\bibitem{ibuka2014free}
\bibinfo{author}{Ibuka, Y.}, \bibinfo{author}{Li, M.}, \bibinfo{author}{Vietri, J.}, \bibinfo{author}{Chapman, G.~B.} \& \bibinfo{author}{Galvani, A.~P.}
\newblock \bibinfo{journal}{\bibinfo{title}{Free-riding behavior in vaccination decisions: an experimental study}}.
\newblock {\emph{\JournalTitle{PloS one}}} \textbf{\bibinfo{volume}{9}}, \bibinfo{pages}{e87164}, \doiprefix\url{https://doi.org/10.1371/journal.pone.0087164} (\bibinfo{year}{2014}).

\bibitem{chapman2012using}
\bibinfo{author}{Chapman, G.~B.} \emph{et~al.}
\newblock \bibinfo{journal}{\bibinfo{title}{Using game theory to examine incentives in influenza vaccination behavior}}.
\newblock {\emph{\JournalTitle{Psychological science}}} \textbf{\bibinfo{volume}{23}}, \bibinfo{pages}{1008--1015}, \doiprefix\url{https://doi.org/10.1177/0956797612437606} (\bibinfo{year}{2012}).

\bibitem{hershey1994roles}
\bibinfo{author}{Hershey, J.~C.}, \bibinfo{author}{Asch, D.~A.}, \bibinfo{author}{Thumasathit, T.}, \bibinfo{author}{Meszaros, J.} \& \bibinfo{author}{Waters, V.~V.}
\newblock \bibinfo{journal}{\bibinfo{title}{The roles of altruism, free riding, and bandwagoning in vaccination decisions}}.
\newblock {\emph{\JournalTitle{Organizational behavior and human decision processes}}} \textbf{\bibinfo{volume}{59}}, \bibinfo{pages}{177--187}, \doiprefix\url{https://doi.org/10.1006/obhd.1994.1055} (\bibinfo{year}{1994}).

\bibitem{abdallah2021social}
\bibinfo{author}{Abdallah, D.~A.} \& \bibinfo{author}{Lee, C.~M.}
\newblock \bibinfo{journal}{\bibinfo{title}{{Social norms and vaccine uptake: College students’ COVID vaccination intentions, attitudes, and estimated peer norms and comparisons with influenza vaccine}}}.
\newblock {\emph{\JournalTitle{Vaccine}}} \textbf{\bibinfo{volume}{39}}, \bibinfo{pages}{2060--2067}, \doiprefix\url{https://doi.org/10.1016/j.vaccine.2021.03.018} (\bibinfo{year}{2021}).

\bibitem{schmelz2021overcoming}
\bibinfo{author}{Schmelz, K.} \& \bibinfo{author}{Bowles, S.}
\newblock \bibinfo{journal}{\bibinfo{title}{{Overcoming COVID-19 vaccination resistance when alternative policies affect the dynamics of conformism, social norms, and crowding out}}}.
\newblock {\emph{\JournalTitle{Proceedings of the National Academy of Sciences}}} \textbf{\bibinfo{volume}{118}}, \bibinfo{pages}{e2104912118}, \doiprefix\url{https://doi.org/10.1073/pnas.2104912118} (\bibinfo{year}{2021}).

\bibitem{betsch2016detrimental}
\bibinfo{author}{Betsch, C.} \& \bibinfo{author}{B{\"o}hm, R.}
\newblock \bibinfo{journal}{\bibinfo{title}{Detrimental effects of introducing partial compulsory vaccination: experimental evidence}}.
\newblock {\emph{\JournalTitle{The European Journal of Public Health}}} \textbf{\bibinfo{volume}{26}}, \bibinfo{pages}{378--381}, \doiprefix\url{https://doi.org/10.1093/eurpub/ckv154} (\bibinfo{year}{2016}).

\bibitem{korn2017drawbacks}
\bibinfo{author}{Korn, L.}, \bibinfo{author}{Betsch, C.}, \bibinfo{author}{B{\"o}hm, R.} \& \bibinfo{author}{Meier, N.~W.}
\newblock \bibinfo{journal}{\bibinfo{title}{Drawbacks of communicating refugee vaccination rates}}.
\newblock {\emph{\JournalTitle{The Lancet Infectious Diseases}}} \textbf{\bibinfo{volume}{17}}, \bibinfo{pages}{364--365}, \doiprefix\url{https://doi.org/10.1016/S1473-3099(17)30141-X} (\bibinfo{year}{2017}).

\bibitem{bohm2017behavioural}
\bibinfo{author}{B{\"o}hm, R.}, \bibinfo{author}{Meier, N.~W.}, \bibinfo{author}{Korn, L.} \& \bibinfo{author}{Betsch, C.}
\newblock \bibinfo{journal}{\bibinfo{title}{Behavioural consequences of vaccination recommendations: an experimental analysis}}.
\newblock {\emph{\JournalTitle{Health economics}}} \textbf{\bibinfo{volume}{26}}, \bibinfo{pages}{66--75}, \doiprefix\url{https://doi.org/10.1002/hec.3584} (\bibinfo{year}{2017}).

\bibitem{korn2018social}
\bibinfo{author}{Korn, L.}, \bibinfo{author}{Betsch, C.}, \bibinfo{author}{B{\"o}hm, R.} \& \bibinfo{author}{Meier, N.~W.}
\newblock \bibinfo{journal}{\bibinfo{title}{{Social nudging: The effect of social feedback interventions on vaccine uptake.}}}
\newblock {\emph{\JournalTitle{Health Psychology}}} \textbf{\bibinfo{volume}{37}}, \bibinfo{pages}{1045}, \doiprefix\url{https://doi.org/10.1037/hea0000668} (\bibinfo{year}{2018}).

\bibitem{bohm2019willingness}
\bibinfo{author}{B{\"o}hm, R.}, \bibinfo{author}{Meier, N.~W.}, \bibinfo{author}{Gro{\ss}, M.}, \bibinfo{author}{Korn, L.} \& \bibinfo{author}{Betsch, C.}
\newblock \bibinfo{journal}{\bibinfo{title}{The willingness to vaccinate increases when vaccination protects others who have low responsibility for not being vaccinated}}.
\newblock {\emph{\JournalTitle{Journal of Behavioral Medicine}}} \textbf{\bibinfo{volume}{42}}, \bibinfo{pages}{381--391}, \doiprefix\url{https://doi.org/10.1007/s10865-018-9985-9} (\bibinfo{year}{2019}).

\bibitem{meier2020individual}
\bibinfo{author}{Meier, N.~W.}, \bibinfo{author}{B{\"o}hm, R.}, \bibinfo{author}{Korn, L.} \& \bibinfo{author}{Betsch, C.}
\newblock \bibinfo{journal}{\bibinfo{title}{Individual preferences for voluntary vs. mandatory vaccination policies: an experimental analysis}}.
\newblock {\emph{\JournalTitle{European Journal of Public Health}}} \textbf{\bibinfo{volume}{30}}, \bibinfo{pages}{50--55}, \doiprefix\url{https://doi.org/10.1093/eurpub/ckz181} (\bibinfo{year}{2020}).

\bibitem{korn2020vaccination}
\bibinfo{author}{Korn, L.}, \bibinfo{author}{Böhm, R.}, \bibinfo{author}{Meier, N.~W.} \& \bibinfo{author}{Betsch, C.}
\newblock \bibinfo{journal}{\bibinfo{title}{Vaccination as a social contract}}.
\newblock {\emph{\JournalTitle{Proceedings of the National Academy of Sciences}}} \textbf{\bibinfo{volume}{117}}, \bibinfo{pages}{14890--14899}, \doiprefix\url{https://doi.org/10.1073/pnas.1919666117} (\bibinfo{year}{2020}).

\bibitem{sheeran2016intention}
\bibinfo{author}{Sheeran, P.} \& \bibinfo{author}{Webb, T.~L.}
\newblock \bibinfo{journal}{\bibinfo{title}{The intention--behavior gap}}.
\newblock {\emph{\JournalTitle{Social and personality psychology compass}}} \textbf{\bibinfo{volume}{10}}, \bibinfo{pages}{503--518}, \doiprefix\url{https://doi.org/10.1111/spc3.12265} (\bibinfo{year}{2016}).

\bibitem{sekara2014strength}
\bibinfo{author}{Sekara, V.} \& \bibinfo{author}{Lehmann, S.}
\newblock \bibinfo{journal}{\bibinfo{title}{The strength of friendship ties in proximity sensor data}}.
\newblock {\emph{\JournalTitle{PloS one}}} \textbf{\bibinfo{volume}{9}}, \bibinfo{pages}{e100915} (\bibinfo{year}{2014}).

\bibitem{stopczynski2014measuring}
\bibinfo{author}{Stopczynski, A.} \emph{et~al.}
\newblock \bibinfo{journal}{\bibinfo{title}{Measuring large-scale social networks with high resolution}}.
\newblock {\emph{\JournalTitle{PloS one}}} \textbf{\bibinfo{volume}{9}}, \bibinfo{pages}{e95978}, \doiprefix\url{https://doi.org/10.1371/journal.pone.0095978} (\bibinfo{year}{2014}).

\bibitem{raudenbush2002hierarchical}
\bibinfo{author}{Raudenbush, S.~W.} \& \bibinfo{author}{Bryk, A.~S.}
\newblock \emph{\bibinfo{title}{{Hierarchical linear models: Applications and data analysis methods}}}, vol.~\bibinfo{volume}{1} (\bibinfo{publisher}{sage}, \bibinfo{year}{2002}).

\bibitem{survival_book}
\bibinfo{author}{Therneau, T.~M.} \& \bibinfo{author}{Grambsch, P.~M.}
\newblock \emph{\bibinfo{title}{{Modeling Survival Data: Extending the Cox Model}}} (\bibinfo{publisher}{Springer}, \bibinfo{address}{New York}, \bibinfo{year}{2000}).

\bibitem{influenza_r0}
\bibinfo{author}{Chowell, G.}, \bibinfo{author}{Miller, M.~A.} \& \bibinfo{author}{Viboud, C.}
\newblock \bibinfo{journal}{\bibinfo{title}{{Seasonal influenza in the United States, France, and Australia: transmission and prospects for control}}}.
\newblock {\emph{\JournalTitle{Epidemiology and Infection}}} \textbf{\bibinfo{volume}{136}}, \bibinfo{pages}{852–864}, \doiprefix\url{https://doi.org/10.1017/s0950268807009144} (\bibinfo{year}{2008}).

\bibitem{nichol1996side}
\bibinfo{author}{Nichol, K.~L.} \emph{et~al.}
\newblock \bibinfo{journal}{\bibinfo{title}{Side effects associated with influenza vaccination in healthy working adults: a randomized, placebo-controlled trial}}.
\newblock {\emph{\JournalTitle{Archives of Internal Medicine}}} \textbf{\bibinfo{volume}{156}}, \bibinfo{pages}{1546--1550}, \doiprefix\url{https://doi.org/10.1001/archinte.1996.00440130090009} (\bibinfo{year}{1996}).

\bibitem{bogner2016response}
\bibinfo{author}{Bogner, K.} \& \bibinfo{author}{Landrock, U.}
\newblock \bibinfo{journal}{\bibinfo{title}{Response biases in standardised surveys}}.
\newblock {\emph{\JournalTitle{GESIS survey guidelines}}} \doiprefix\url{https://doi.org/10.15465/gesis-sg_en_016} (\bibinfo{year}{2016}).

\bibitem{podsakoff2003common}
\bibinfo{author}{Podsakoff, P.~M.}, \bibinfo{author}{MacKenzie, S.~B.}, \bibinfo{author}{Lee, J.-Y.} \& \bibinfo{author}{Podsakoff, N.~P.}
\newblock \bibinfo{journal}{\bibinfo{title}{Common method biases in behavioral research: a critical review of the literature and recommended remedies.}}
\newblock {\emph{\JournalTitle{Journal of applied psychology}}} \textbf{\bibinfo{volume}{88}}, \bibinfo{pages}{879}, \doiprefix\url{https://doi.org/10.1037/0021-9010.88.5.879} (\bibinfo{year}{2003}).

\bibitem{stopczynski2014privacy}
\bibinfo{author}{Stopczynski, A.}, \bibinfo{author}{Pietri, R.}, \bibinfo{author}{Pentland, A.}, \bibinfo{author}{Lazer, D.} \& \bibinfo{author}{Lehmann, S.}
\newblock \bibinfo{journal}{\bibinfo{title}{{Privacy in sensor-driven human data collection: A guide for practitioners}}}.
\newblock {\emph{\JournalTitle{arXiv preprint arXiv:1403.5299}}} \doiprefix\url{https://doi.org/10.48550/arXiv.1403.5299} (\bibinfo{year}{2014}).

\bibitem{sapiezynski2015tracking}
\bibinfo{author}{Sapiezynski, P.}, \bibinfo{author}{Stopczynski, A.}, \bibinfo{author}{Gatej, R.} \& \bibinfo{author}{Lehmann, S.}
\newblock \bibinfo{journal}{\bibinfo{title}{Tracking human mobility using wifi signals}}.
\newblock {\emph{\JournalTitle{PloS one}}} \textbf{\bibinfo{volume}{10}}, \bibinfo{pages}{e0130824}, \doiprefix\url{https://doi.org/10.1371/journal.pone.0130824} (\bibinfo{year}{2015}).

\end{thebibliography}

\end{document}